\documentclass[%
preprintnumbers,
showkeys,
amsmath,amssymb,
aps,
]{revtex4-2}

\usepackage{graphicx}% Include figure files
\usepackage{dcolumn}% Align table columns on decimal point
\usepackage{bm}% bold math
\usepackage{subcaption, float}
\newcommand{\pder}[2]{\frac{\partial#1}{\partial#2}} % partial derivative
\newcommand{\tder}[2]{\frac{\mathrm{d}#1}{\mathrm{d}#2}} % partial derivative
\newcommand{\tderi}[2]{\mathrm{d}#1 / \mathrm{d}#2} % partial derivative inline fraction for text
\begin{document}

\preprint{APS/123-PRF}

%\title{Manuscript Title:\\with Forced Linebreak}% Force line breaks with \\
%\title{A new ODE-based wall model for boundary layers accounting for pressure gradient and Reynolds number effects}
\title{A new ODE-based turbulence wall model accounting for pressure gradient and Reynolds number effects}

\author{Kevin Patrick Griffin}
 \email{kevinpg@stanford.edu}
\author{Lin Fu}
\affiliation{%
Center for Turbulence Research, Stanford University, Stanford, CA 94305-3024, United States of America
}%

\date{\today}% It is always \today, today,
\begin{abstract}
In wall-modeled large-eddy simulations (WMLES), the near-wall model plays a significant role in predicting the skin friction, although the majority of the boundary layer is resolved by the outer large-eddy simulation (LES) solver.
In this work, we aim at developing a new ordinary differential equation (ODE)-based wall model, which is as simple as the classical equilibrium model yet capable of capturing non-equilibrium effects and low Reynolds number effects. The proposed model reformulates the classical equilibrium model by introducing a new non-dimensional mixing-length function. The new mixing-length function is parameterized in terms of the boundary layer shape factor instead of the commonly used pressure-gradient parameters. As a result, the newly introduced mixing-length function exhibits great universality within the viscous sublayer, the buffer layer, and the log region (i.e., $0 < y < 0.1\delta$, where the wall model is typically deployed in a WMLES setup). The performance of the new model is validated by predicting a wide range of canonical flows with the friction Reynolds number between 200 and 5200, and the Clauser pressure-gradient parameter between -0.3 and 4. Compared to the classical equilibrium wall model, remarkable error reduction in terms of the skin friction prediction is obtained by the new model. Moreover, since the new model is ODE-based, it is straightforward to be deployed for predicting flows with complex geometries and therefore promising for a wide range of applications.
\end{abstract}

\maketitle

%\tableofcontents

%%%%%%%%%%%%%%%%%%%%%%%%%%%%%%%%%%%%%%%%%%%%%%%%%%%%%%%%%%%%%%
\section{\label{sec:intro}Introduction}
%%%%%%%%%%%%%%%%%%%%%%%%%%%%%%%%%%%%%%%%%%%%%%%%%%%%%%%%%%%%%%

The accurate prediction of wall-bounded turbulence is practically important for many engineering applications, e.g. the design of low-drag vehicles. While direct numerical simulation (DNS) and LES are capable of delivering accurate solutions, the required number of grid points scales with 
%$Re^{2.05}$ \cite{Yang2020}
$Re^{2.64}$
and $Re^{1.86}$ \cite{choi2012grid} respectively, rendering them prohibitively expensive for high-Reynolds-number flows. Alternatively, the computational cost of the WMLES approach, which resolves the large-scale energetic motion in the outer portion of the boundary layer and employs a reduced-order model for the near-wall turbulence, scales only linearly with Reynolds number \cite{choi2012grid}. Due to the fact that the LES subgrid-scale model itself typically provides an inconsistent wall shear stress when the near-wall turbulence eddies are poorly resolved, the performance of WMLES heavily relies on the wall model. The main wall models include the wall-stress-based models \cite{deardorff1970numerical,bose2018wall,larsson2016large}, the detached-eddy simulation (DES) paradigm \cite{spalart2009detached}, the dynamic slip wall model \cite{bose2014dynamic,bae2019dynamic,Griffin2018,Griffin2019}, integral-based models (e.g. \cite{Yang2015}), and other variants. In this work, our discussions are restricted to the wall-stress-based models, and the readers are referred to \cite{piomelli2002wall,sagaut2006large,bose2018wall} for more comprehensive reviews on other wall models.

The core idea of wall-stress-based models is to develop a computationally efficient reduced-order model (e.g. the Reynolds-averaged Navier-Stokes (RANS)-like models) such that the physically correct wall shear stress can be estimated by solving this model between the wall and the matching location, where the instantaneous LES data is provided to the wall model, and fed back to the LES solver in the outer boundary layer through a wall boundary condition. 

Most ODE-based wall models are only strictly valid for equilibrium flows, which feature a constant edge condition, such as the fully-developed pipe or channel flows, and the zero pressure gradient boundary layer flows. Conversely, a non-equilibrium flow is characterized by an edge condition that is non-constant and evolves spatially. To simulate such a flow, the natural choice is to solve the full boundary layer partial differential equations (PDEs) as a wall-stress-based model such that some non-equilibrium effects can be captured \cite{balaras1996two,park2014improved,wang2002dynamic,kawai2013dynamic}. The main disadvantages of solving the boundary layer PDEs are the significantly increased computational cost and the requirement of a high-quality near-wall mesh, which is non-trivial to generate for complex geometries.
By neglecting the temporal term, the convective flux and the pressure-gradient term in the turbulent boundary layer RANS equations, the equilibrium wall model has been proposed based on the classical mixing-length eddy viscosity model. The equilibrium wall model has become increasingly popular for engineering applications since only ODEs are solved in the wall-normal direction instead of the more expensive PDEs as in the classical RANS models, e.g. see \cite{bermejo2014confinement,fu2020heat,Fu2018Equilibrium,iyer2019analysis,mettu2018wall,goc2020wall,boukharfane2020characterization}. To further improve the predictive capability, several variants of non-equilibrium wall models have also been proposed by retaining part of the neglected terms, e.g. \cite{hoffmann1995approximate,wang2002dynamic,catalano2003numerical,duprat2011wall,chen2014wall}. However, Hickel et al. \cite{hickel2013parametrized} claim that these neglected terms should appear together if any of them is retained since they balance each other (this conclusion will be challenged in section~\ref{sec:eqwm_challenges} of the present work).

In this work, the limitations of existing ODE-based wall models are analyzed and a new ODE-based inner wall model is proposed accounting for pressure gradient and Reynolds number effects. The classical mixing-length-based equilibrium wall model is reformulated without appealing to the existence of the constant-shear-stress layer. Moreover, instead of relying on pressure-gradient parameters, the new model sensitizes the law of the wall to the boundary layer shape factor, which can be robustly computed with the information from the outer solver and the inner wall model. As a result, the proposed wall model greatly extends the predictive capability of the wall model for flows with strong pressure gradients and a wide range of Reynolds numbers. Note that, the inner wall model proposed in this work is suitable for deployment with various outer PDE solvers, e.g. the RANS equations or the LES equations, although most of following discussions are restricted to the context of the WMLES paradigm.

The remaining of this paper is organized as follows. (i) In section \uppercase\expandafter{\romannumeral2}, the classical equilibrium ODE-based wall model is reviewed and the corresponding limitations are analyzed. (ii) In section \uppercase\expandafter{\romannumeral3}, the new model accounting for pressure gradient and Re effects is developed. (iii) In section \uppercase\expandafter{\romannumeral4}, the performance of the proposed model is validated for predicting a wide range of flows. (iv) In section \uppercase\expandafter{\romannumeral5}, concluding discussions and remarks are given.

%%%%%%%%%%%%%%%%%%%%%%%%%%%%%%%%%%%%%%%%%%%%%%%%%%%%%%%%%%%%%%
\section{\label{sec:Equilibrium_wall_model} Classical equilibrium ODE-based wall model}
%%%%%%%%%%%%%%%%%%%%%%%%%%%%%%%%%%%%%%%%%%%%%%%%%%%%%%%%%%%%%%

The classical ODE-based wall model takes the definition of the total shear stress $\tau$ as the starting point, i.e.
\begin{equation} \label{eq:define_tau_dim}
    \frac{\tau}{\rho} = (\nu + \nu_t) \pder{U}{y},
\end{equation}
where $y$ is the wall-normal coordinate, $U$ is the mean streamwise velocity, $\rho$ is the fluid density, $\nu$ is the kinematic viscosity, and $\nu_t$ is the modeled turbulent eddy viscosity. To render this equation an ODE, it is assumed that the mean streamwise velocity $U$ profile is only a function of $y$, i.e.
\begin{equation} \label{eq:dudy_dim}
    \tder{U}{y} = \frac{\tau/\rho}{\nu + \nu_t}.
\end{equation}
To close this ODE for $U[y]$, the classical approach is to define $\tau[y]$ and $\nu_t[y]$ based on the equilibrium assumption and the mixing-length model, respectively. Typically the so-called constant-stress-layer assumption is invoked to assume that $\tau = \tau_w$ is a constant over the domain where the wall model will be deployed. The rationality of this assumption relies on the fact that the temporal, convective, and pressure-gradient terms of the turbulent boundary layer equations approximately balance each other in the near-wall region when the flow Reynolds number is sufficiently high \cite{hickel2013parametrized}. The remaining modeling issue is to develop a consistent model to parameterize the $\nu_t[y]$ profile.

%%%%%%%%%%%%%%%%%%%%%%%%%%%%%%%%%%%%%%%%%%%%%%
\subsection{Classical models for the eddy viscosity}
%%%%%%%%%%%%%%%%%%%%%%%%%%%%%%%%%%%%%%%%%%%%%%

While there are more sophisticated and also probably more general one-equation \cite{spalart1992one} and two-equation \cite{menter1994two} RANS models for $\nu_t[y]$, the present work focuses on the ``zero-equation" models for the simple and efficient deployment into the WMLES framework without requiring solving PDEs on a separate tailored mesh.

There are two classical ``zero-equation" models for the eddy viscosity $\nu_t[y]$. Both are derived using dimensional arguments to predict the eddy viscosity from a mixing length scale $\ell$ and either a mixing time scale or a mixing velocity scale. The former is called Prandtl's mixing-length model \cite{Prandtl1925}, and 
\begin{equation}
    \nu_t =  \ell_{P}^2 \left| \tder{U}{y} \right|,
    \label{eq:pr_model}
\end{equation}
where $\ell_P$ denotes Prandtl's mixing length, and the inverse of the mean shear denotes Prandtl's mixing time scale.

Alternatively, Cabot \cite{Cabot1995} defines the eddy viscosity as 
\begin{equation}
    \nu_t = \ell_{C} u_\tau ,
    \label{eq:cabot_model}
\end{equation}
where the wall friction velocity $u_\tau = \sqrt{\tau_w/\rho}$ and the wall shear stress $\tau_w = \tau|_{y=0}$. 
Note that, often this model is attributed to Johnson and King \cite{Johnson1985}, but their model uses the square root of the peak of the Reynolds shear stress as the velocity scale instead of $u_\tau$.

To deploy either of these two models to integrate the nonlinear ODE, i.e. Eq.~(\ref{eq:dudy_dim}), the length-scale $\ell[y]$ must be specified in terms of the wall-normal distance. van Driest \cite{VANDRIEST1956} shows that a logarithmic velocity profile (the so-called log law) with a specific logarithmic intercept constant can be recovered by invoking the constant-stress-layer assumption and letting
\begin{equation} \label{eq:damped_mixing_length}
    \ell = \kappa y D[y],
\end{equation}
where $\kappa$ denotes the Kármán constant, and $D$ is a damping function that approaches unity for large values of $y$ and zero for vanishing $y$. Plugging van Driest's mixing length (Eq.~(\ref{eq:damped_mixing_length})) into either Prandtl's or Cabot's eddy viscosity model and then substituting the resultant $\nu_t$ into Eq.~(\ref{eq:dudy_dim}) leads to the following relation
\begin{equation} \label{eq:dudy_log_law}
    \tder{U}{y} = \frac{u_\tau}{\kappa y},
\end{equation}
in the limit of large $y$. Throughout the following discussions, the superscript $+$ refers to quantities non-dimensionalized via the viscous length scale $\delta_v = \nu/u_\tau$ and the velocity scale $u_\tau$. By non-dimensionalizing and integrating Eq.~(\ref{eq:dudy_log_law}), the log law is obtained as
\begin{equation}
    U^+ = \frac{1}{\kappa} \ln(y^+) + B,
\end{equation}
where the parameter $1/\kappa$ denotes the log slope and B denotes the log intercept constant. van Driest \cite{VANDRIEST1956} shows that the near-wall behavior in the viscous sub-layer and buffer layer can also be asymptotically recovered by carefully choosing $D[y]$, and this choice implicitly determines the log intercept constant $B$.

\begin{figure}
  \centering
  \includegraphics[width=0.5\linewidth]{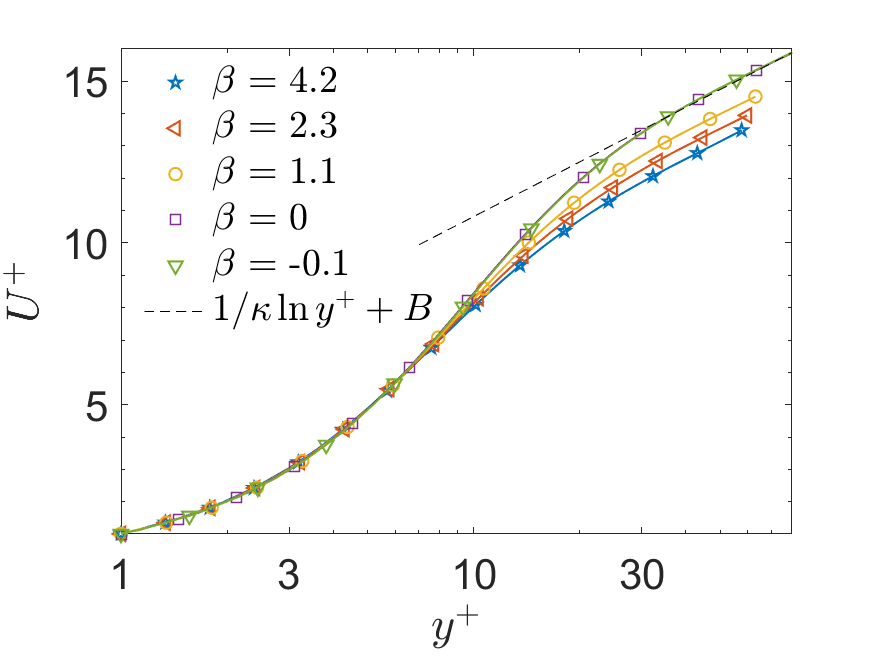}
  \caption{The distribution of the mean streamwise velocity $U^+$ plotted versus the wall-normal coordinate $y^+$ for five cases from Table~\ref{tab:database}. Included are three APGBLs ($\beta = 4.2, 2.3, 1.1$), a ZPGBL at $Re_\tau = 2000$ ($\beta = 0$), and a channel flow at $Re_\tau = 5200$ ($\beta = -0.1$). Also plotted is the reference log law distribution with the Kármán constant $\kappa = 0.41$ and the intercept constant $B$ = 5.2. All data is truncated at $y=0.1\delta$, which approximately encompasses the viscous sublayer, buffer layer, and log layer.}
  \label{fig:u_vs_y}
\end{figure}

For near-equilibrium flows, Cabot's and Prandtl's models are quite successful since the log law coefficients $\kappa$ and $B$ are relatively robust for flows with a wide range of Reynolds numbers. However, for flows where non-equilibrium effects are significant, such as the adverse pressure gradient boundary layers shown in Fig.~\ref{fig:u_vs_y}, the log intercept constant $B$ exhibits non-universality. The log slope also exhibits non-universality but to a lesser extent. Therefore, by construction, Prandtl's and Cabot's models will predict the same value of $\kappa$ and $B$ regardless of the outer flows, thus these models will fail in non-equilibrium flows. As will be discussed in section~\ref{sec:results}, for the adverse pressure gradient boundary layer, deployment of Cabot's model can lead to a $17\%$ under-prediction of the wall shear stress. Similarly, $B$, and to a lesser extent $\kappa$, vary in flows with rather low Reynolds numbers.

\begin{table}[]
\begin{tabular}{lcccccc}
\hline
\multicolumn{1}{c}{Type of flow} & $Re_\tau$ & $1000\alpha$ & $\beta$ & $H$ & Number of profiles & Sources \\ \hline
ZPGBLs & 	[276,2479]	 & [ 0.00, 0.00] & [ 0.00, 0.00] & [1.36,1.49] & 8 & \cite{Sillero2013,Spalart1988,Eitel-Amor2014} \\
Channels & 	[543,5186]	 & [-1.84,-0.19] & [-0.13,-0.10] & [1.25,1.40] & 4 & \cite{Lee2015} \\
Pipes & 	[685,1143]	 & [-2.92,-1.75] & [-0.28,-0.28] & [1.37,1.40] & 2 & \cite{Wu2008} \\
APGBLs & 	[202, 740]	 & [ 5.01,31.30] & [ 0.19, 4.43] & [1.56,1.91] & 21 & \cite{Bobke2017} \\
NACA 0012 AoA $=0^{\circ}$ & 	[264, 371]	 & [ 5.36,18.83] & [ 0.30, 1.71] & [1.55,1.59] & 3 & \cite{Tanarro2020} \\
NACA 4412 AoA $=5^{\circ}$ & 	[290, 679]	 & [-1.32,12.50] & [-0.14, 2.08] & [1.41,1.59] & 8 & \cite{Vinuesa2018} \\
\hline
\end{tabular}
\caption{Well-resolved simulation database from various types of flows for wall model evaluation. Included are zero pressure gradient boundary layers (ZPGBLs), fully-developed channel and pipe flows, adverse pressure gradient boundary layers (APGBLs) with five different pressure gradient conditions, and two airfoil flows with specified angle of attack (AoA). The ranges of friction Reynolds number $Re_\tau$, the inner pressure-gradient parameter $\alpha =  (\delta_\nu /\tau_w) \tderi{P}{x}$, the outer pressure-gradient parameter $\beta = (\delta^* / \tau_w) \tderi{P}{x}$, and the boundary layer shape factor $H$ are also provided, as well as the number of 1D profiles available for each flow.}
\label{tab:database}
\end{table}
%%

%%%%%%%%%%%%%%%%%%%%%%%%%%%%%%%%%%%%%%%%%%%%%%
\subsection{Challenges of extending the classical models to non-equilibrium flows} \label{sec:eqwm_challenges}
%%%%%%%%%%%%%%%%%%%%%%%%%%%%%%%%%%%%%%%%%%%%%%

\begin{figure}
  \centering
  \includegraphics[width=0.5\linewidth]{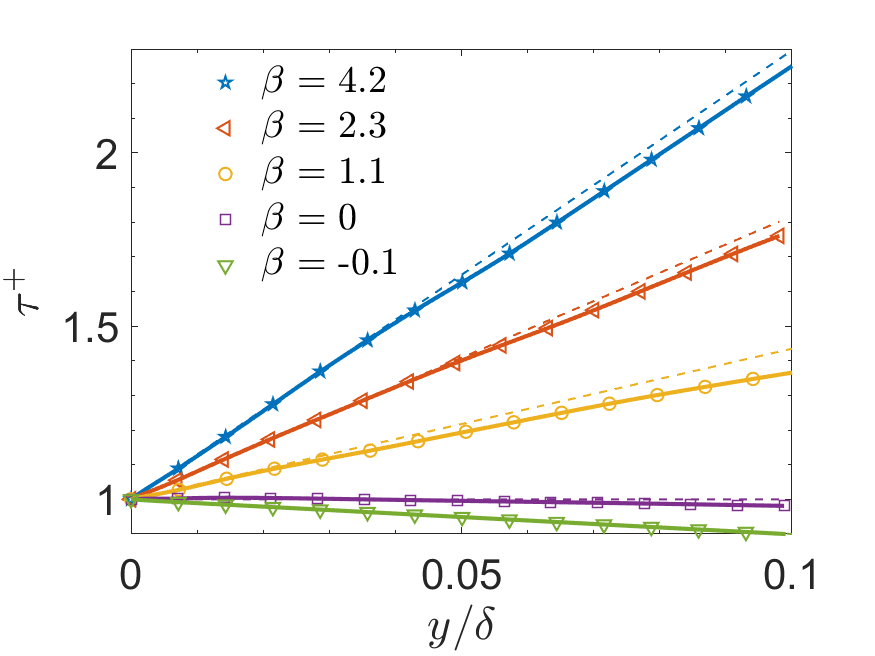}
  \caption{The total shear stress $\tau^+$ profiles (solid lines) versus the wall-normal coordinate normalized by the boundary layer thickness $\delta$, for the same cases as shown in Fig.~\ref{fig:u_vs_y}. Also given are the linear shear stress approximations (dashed lines) predicted by Eq.~(\ref{eq:linear_tau}). All data is truncated at $y=0.1\delta$, which approximately encompasses the viscous sublayer, buffer layer, and log layer.}
  \label{fig:tau_vs_y}
\end{figure}

\begin{figure}
  \centering
  \includegraphics[width=0.5\linewidth]{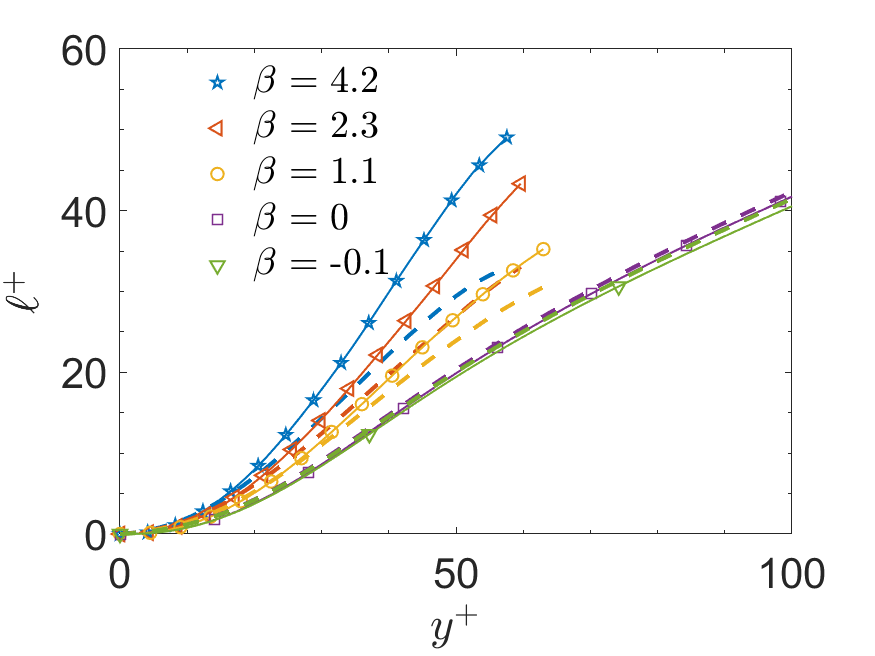}
  \caption{Distributions of Cabot's mixing length $\ell_C^+$ (solid lines) and Prandtl's mixing length $\ell_P^+$ (dashed lines) plotted versus the wall-normal coordinate $y^+$, for the same cases as shown in Fig.~\ref{fig:u_vs_y}. All data is truncated at $y=0.1\delta$, which approximately encompasses the viscous sublayer, buffer layer, and log layer.}
   \label{fig:ell_vs_y}
\end{figure}

The basis of the equilibrium model is that there is a universal mixing-length profile (either Cabot's or Prandtl's) and a constant total-shear-stress layer in the near-wall region. However, neither of these assumptions are valid for non-equilibrium flows. As shown in Fig.~\ref{fig:tau_vs_y}, for an adverse pressure gradient boundary layer, the total shear stress varies by up to $125\%$ in the wall-modeled region. Similarly, Cabot's and Prandtl's mixing-length profiles vary from case to case by $50\% - 100\%$, see Fig.~\ref{fig:ell_vs_y}. 

Since Cabot's model assumes a constant stress layer and a universal $\nu_t$ profile (fitted from canonical equilibrium flows), the wrong stress and eddy viscosity profiles are fed to the modeling ODE, i.e. Eq.~(\ref{eq:dudy_dim}). However, it will, by construction, recover the log law as analyzed in the previous section. For boundary layer flows with modest departure from the log law, as shown in Fig.~\ref{fig:u_vs_y}, Cabot's (or Prandtl's) model will generate a smaller error than one might expect by examining the model inputs, i.e. the $\tau$ or $\nu_t$ profiles. The errors in these inputs cancel each other due to the fact that only the ratio of these terms appears in the modeling ODE (Eq.~(\ref{eq:dudy_dim})).

\begin{figure}
  \centering
  \includegraphics[width=0.5\linewidth]{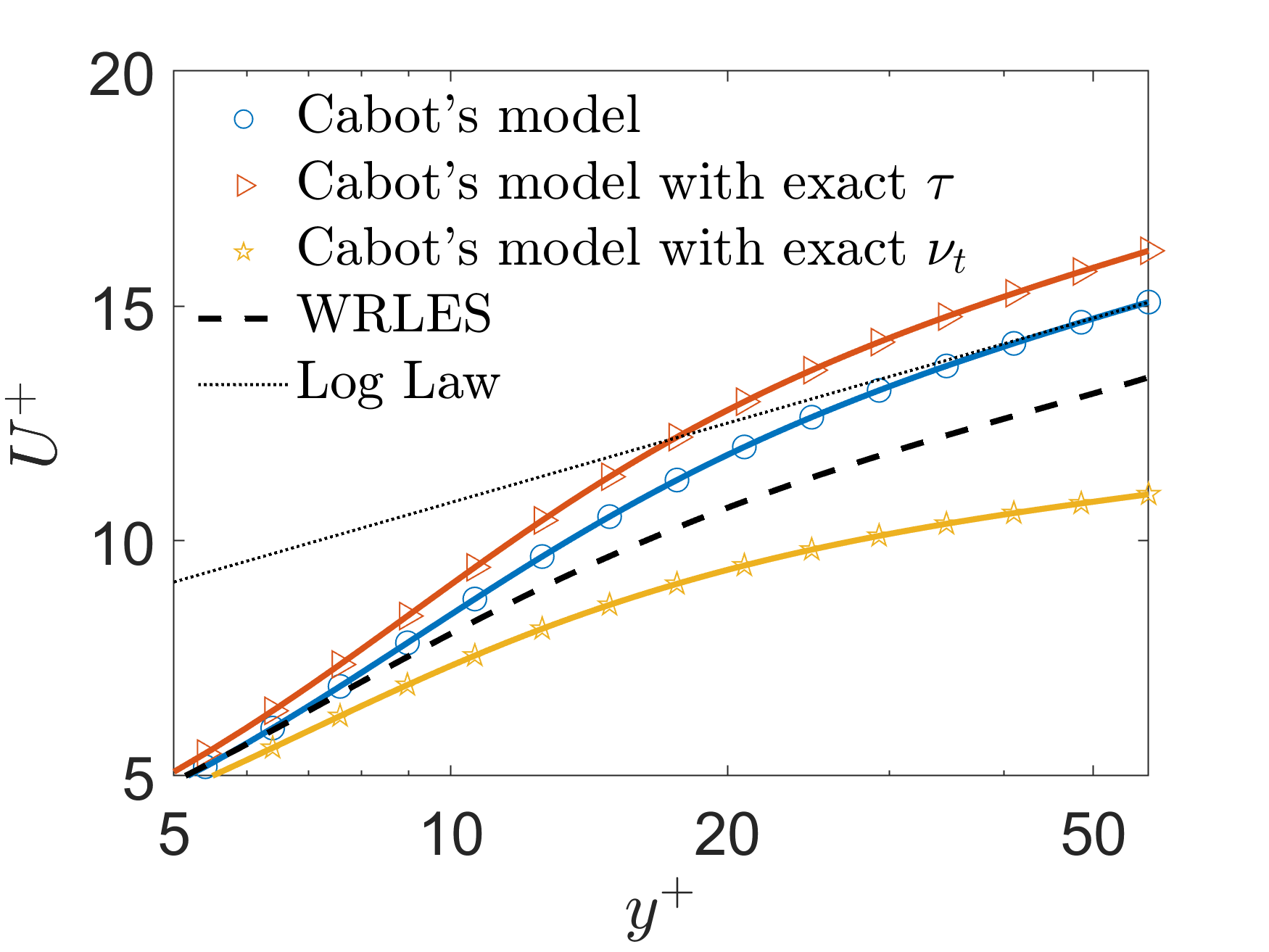}
  \caption{The velocity profile $U^+$ plotted versus the wall-normal coordinate $y^+$, for the $\beta=4.2$ case considered in Fig.~\ref{fig:u_vs_y}. Included are the WRLES data (dashed line), which suggests a shift with respect to the log law reference (dotted line), and the three predictions from variants of Cabot's model (symbols). 
  }
   \label{fig:LOW_u_vs_y_old_model}
\end{figure}

This carefully-designed error cancellation enhances the predictive capability of the equilibrium models and partially explains why these models perform decently in non-equilibrium flows, such as in flows over a swept wing \cite{goc2020wall}. However, the error cancellation makes it more complicated to identify and remove the remaining errors that do not cancel. Even if the $\nu_t$ or $\tau$ models are improved independently, there is no guarantee that the prediction from the resulting ODE model agrees better with the exact solution $U[y]$. This paradox is demonstrated in Fig.~\ref{fig:LOW_u_vs_y_old_model}. In a non-equilibrium setting, Cabot's model fails to correctly predict the log law intercept constant $B$. When the exact stress profile (instead of assuming a constant stress layer) is provided, the prediction counter-intuitively becomes worse. Similarly, when the exact eddy viscosity profile is fed 
into the ODE model, the prediction gets worse as well. This is because the mixing-length profile in these models is constructed to recover the log law only in a constant stress layer. When either the $\nu_t$ or the $\tau$ profile is adapted (even improved), the log law is no longer guaranteed with these models. On the other hand, if the model is substituted with both the exact $\nu_t$ and $\tau$ profiles, then the exact $U$ profile would be recovered since Eq.~(\ref{eq:define_tau_dim}) is inherently a definition instead of a model in that case.

Over the past decades, many efforts (e.g. \cite{Galbraith1977,Granville1989,Thomas1989,Bernard2003}) have attempted to include pressure gradient effects in the classical wall model by replacing the constant stress layer assumption with
\begin{equation} \label{eq:linear_tau}
    \tau \approx \tau_w + y \tder{P}{x} = \tau_w (1 + \alpha y^+),
\end{equation}
where the linear term results from retaining the pressure gradient term in the streamwise momentum equation and integrating in the wall-normal direction. On the other hand, Hickel et al. \cite{hickel2013parametrized} argue that three terms of the turbulent boundary layer equations, i.e. the temporal term, the convective flux and the pressure gradient, balance each other and should appear together if any of them is retained. Only including the pressure-gradient term leads to a less accurate description of $\tau[y]$. However, as shown in Fig.~\ref{fig:tau_vs_y}, it is clear that Eq.~(\ref{eq:linear_tau}) is a better model than the constant stress layer and the prediction is in good agreement with the wall-resolved LES (WRLES) data. The fact that only including the pressure-gradient term leads to a worse velocity solution is because the investigation \cite{hickel2013parametrized} has employed the classical mixing-length model, which is designed to recover a log law only when deployed with a constant stress layer. Although a better $\tau$ model is used, a worse prediction of the streamwise-velocity profile $U$ is obtained. This is consistent with the demonstration in Fig.~\ref{fig:LOW_u_vs_y_old_model} that even when an exact $\tau$ model is used, a worse prediction of the $U$ profile is obtained, unless the $\nu_t$ profile is modified accordingly.

Galbraith et al. \cite{Galbraith1977} propose to improve the $\tau$ model and consequently modify the $\nu_t$ model so that the log law is retained. The readers are referred to Appendix~\ref{app:galbriath} for the details of this model. The model is hardwired to always recover a particular log law slope similar to the classical model. One potential advantage of this model is that the resulting log intercept constant depends on the pressure gradient in the flow. However, this turns out to be a weakness of this model since, as remarked in Appendix~\ref{app:galbriath}, the trend is opposite to that observed in high-fidelity WRLES simulations.

Meanwhile, more recently, Meneveau \cite{Meneveau2020} proposes a method which incorporates the pressure gradient without modifying the eddy viscosity profile. This means that the method does not preserve the log law and will be inaccurate in the presence of strong pressure gradients.

In this work, by introducing a novel non-dimensional mixing-length function, a new ODE-based model will be developed to bypass the issue of error cancellation between $\tau$ and $\nu_t$ profiles, and to account for the pressure gradient and low Reynolds number effects more directly.

%%

%%%%%%%%%%%%%%%%%%%%%%%%%%%%%%%%%%%%%%%%%%%%%%%%%%%%%%%%%%%%%%
\section{\label{sec:state_of_the_art} New ODE-based wall model}
%%%%%%%%%%%%%%%%%%%%%%%%%%%%%%%%%%%%%%%%%%%%%%%%%%%%%%%%%%%%%%

The ODE-based model equation, i.e. Eq.~(\ref{eq:dudy_dim}), can be non-dimensionalized using the inner units $u_\tau$ and $\delta_v$, resulting in
\begin{equation} \label{eq:dudy_nondim}
    \tder{U^+}{y^+} = \frac{\tau^+}{1 + \nu_t^+}.
\end{equation}
Given the constant-stress-layer assumption with $\tau^+ = \tau/\tau_w \approx 1$ and that the non-dimensional form of Cabot's model is $\nu_t^+ \approx \ell_C^+$, the Cabot's model ODE is obtained as
\begin{equation}
     \tder{U^+}{y^+} \approx \frac{1}{1 + \ell_{C}^+ }.
    \label{eq:csl_model}
\end{equation}

In this work, in order to avoid the modeling of the stress and the eddy viscosity profile independently, the mean shear is directly parameterized as
\begin{equation}  \label{eq:new_model_ode}
     \tder{U^+}{y^+} = \frac{1}{1+\ell_n^+},
\end{equation}
where $\ell_n^+$ is introduced as an empirical function related to the non-dimensional mean shear and is defined as
\begin{equation}
    \ell_{n}^+ = \frac{1 - \tau^+ + \nu_t^+}{\tau^+}.
    \label{eq:new_ell_m}
\end{equation}
Unlike Cabot's model ODE (i.e. Eq.~(\ref{eq:csl_model})), the new model ODE (i.e. Eq.~(\ref{eq:new_model_ode})) is exact with the above definition of $\ell_n^+$.

By analogy to Cabot's model, $\ell_n^+$ is referred to as a non-dimensional mixing ``length." The new mixing length is equivalent to Cabot's in the special case of a constant stress layer, i.e. $\tau^+ \approx 1$. By rearranging Eq.~(\ref{eq:new_ell_m}), the new mixing length can be viewed as part of a new model for the eddy viscosity
\begin{equation}
    \nu_t^+ = \ell_n^+ \tau^+ + \tau^+ -1.
    %\label{eq:new_model}
\end{equation}
Substituting this model into Eq.~(\ref{eq:dudy_nondim}) leads directly to the new model ODE (i.e. Eq.~(\ref{eq:new_model_ode})).
This model is equivalent in form to the original equilibrium model, but it is exact, even in flows with arbitrary shear stress profiles. 

\begin{figure}
  \centering
  \includegraphics[width=0.5\linewidth]{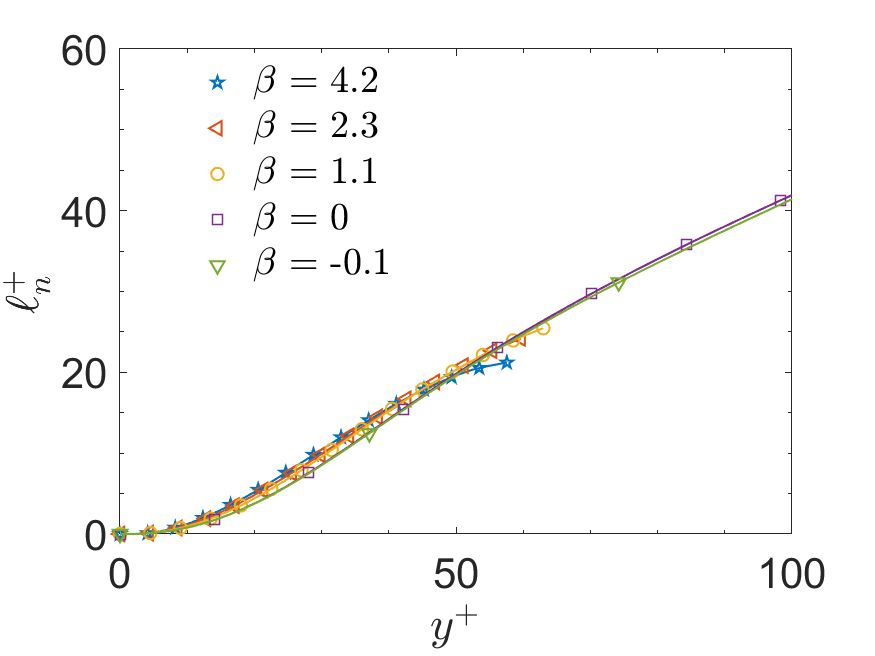}
  \caption{Distributions of the new mixing length $\ell_n^+$ (defined by Eq.~(\ref{eq:new_ell_m})) plotted versus the wall-normal coordinate $y^+$, for the same cases as shown in Fig.~\ref{fig:u_vs_y}. All data is truncated at $y=0.1\delta$, which approximately encompasses the viscous sublayer, buffer layer, and log layer.}
   \label{fig:ell_vs_y_new}
\end{figure}

As shown in Fig.~\ref{fig:ell_vs_y_new}, the profiles of the new mixing lengths collapse well for flows with a wide range of pressure-gradient parameters, whereas they scatter significantly with the classical definitions of Prandtl and Cabot as is indicated by Fig.~\ref{fig:ell_vs_y}. The collapse of the new mixing lengths also indicates the robustness of the non-dimensional mean shear to pressure gradients.

\begin{figure}
\begin{subfigure}{.5\textwidth}
  \centering
  \includegraphics[width=1\linewidth]{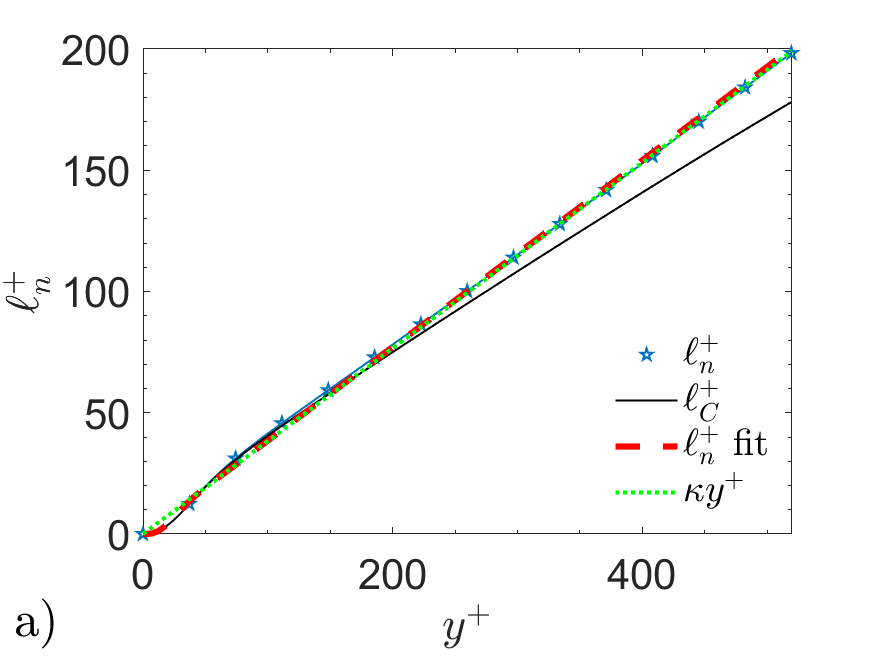}
  \captionlistentry{}
  \label{fig:fit_ell}
\end{subfigure}%
\begin{subfigure}{.5\textwidth}
  \centering
  \includegraphics[width=1\linewidth]{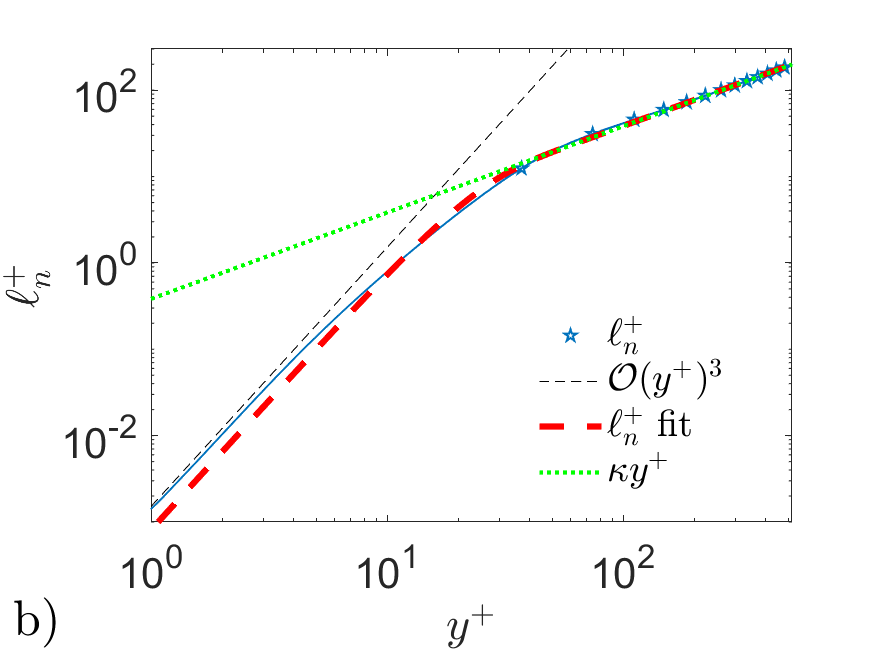}
  \captionlistentry{}
  \label{fig:fit_ell_log}
\end{subfigure}%
\caption{Distributions of the new mixing length $\ell_n^+$ extracted from the DNS database of a channel flow at $Re_\tau \approx 5200$ \cite{Lee2015} plotted versus the wall-normal coordinate $y^+$. In both panels, the log-law relation $\ell^+ = ky^+$ is plotted. Also shown are the distributions of Cabot's mixing length $\ell_C^+$ with linear scale in panel (a), and the $\mathcal{O}(y^+)^3$ reference with log scale in panel (b).}
\label{fig:fits}
\end{figure}

Although Cabot's ODE (i.e. Eq.~(\ref{eq:csl_model})) is equivalent to the exact ODE (i.e. Eq.~(\ref{eq:dudy_nondim})) in the special case that $\tau^+ \approx 1$, in almost all real flows, $\tau^+ \ne 1$. Even if $\ell_C^+=\nu_t^+$ is computed from DNS data and fed into Cabot's ODE, the wrong velocity profile will result. On the other hand, if $\ell_{n}^+ = (1 - \tau^+ + \nu_t^+) / \tau^+$ is computed from DNS data and fed into the new model ODE (i.e. Eq.~(\ref{eq:new_model_ode})), the exact solution will result, by construction. 
For both Cabot's ODE and the new ODE, the mixing lengths should approach a value of $\kappa y^+$ in order to recover the log law with a log slope of $1/\kappa$. When Cabot's mixing length is evaluated from DNS data of a channel flow, $\ell_C^+=\nu_t^+$ does not recover this behavior. As shown in Fig.~\ref{fig:fit_ell}, $\ell_C^+$ deviates from $\kappa y^+$ by $10\%$ in a channel flow at $10\%$ of the channel half height because the total shear stress reduces by $10\%$ at this location.

%%

%%%%%%%%%%%%%%%%%%%%%%%%%%%%%%%%%%%%%%%%%%%%%%%%%%%%%%%%%%%%%%
\subsection{\label{sec:damping} Parameterizing the new mixing length}
%%%%%%%%%%%%%%%%%%%%%%%%%%%%%%%%%%%%%%%%%%%%%%%%%%%%%%%%%%%%%%

By comparing Fig.~\ref{fig:ell_vs_y} with Fig.~\ref{fig:ell_vs_y_new}, it is clear that the new mixing-length definition collapses much better in various flows than the classical definition. In this section, we proceed by determining a suitable functional representation of this new mixing length.

In order to be consistent with physical constraints, the functional representation of the mixing length must recover the well-established asymptotic behavior of the the eddy viscosity near a wall. Concerning the limit of large wall-normal distance, Eq.~(\ref{eq:new_model_ode}) implies that the condition $\ell_n^+ \rightarrow \kappa y^+$ should be satisfied to recover the log-law slope observed in high-Reynolds-number wall-bounded turbulence. 
On the other hand, considering the near-wall region with $y^+\sim \mathcal{O}(1)$, mass conservation and the viscous wall boundary conditions imply that $\nu_t^+ \sim (y^+)^3$ should be satisfied.
%%as asserted by Pope \cite{pope2001} (section 7.1.6).

To examine the effects of these constraints on the mixing length, $\tau$ is expressed as a power series expansion in terms of $\alpha y^+$, where $\alpha$ is the inner pressure-gradient parameter defined in Eq.~(\ref{eq:define_alpha}). Assuming that the pressure gradient is independent of the wall-normal distance $y$, the leading-order term is given by Eq.~(\ref{eq:linear_tau}) after integrating the streamwise momentum equation, and consequently
\begin{equation}
    \tau = 1 + \alpha y^+ + \mathcal{O}(\alpha y^+)^2.
\end{equation}
By plugging the above expansion into Eq.~(\ref{eq:new_ell_m}) and retaining the leading-order terms of $y^+$ and $\alpha y^+$, the expression
\begin{equation}
     \ell_n^+ \sim \mathcal{O}(y^+)^3 + \mathcal{O}(\alpha y^+)
\end{equation}
holds.
According to Table~\ref{tab:database}, $\alpha y^+ << (y^+)^3$ when $y^+\sim \mathcal{O}(1)$, and thus $\ell_n^+ \sim (y^+)^3$ applies in the near-wall region.

As shown in Fig.~\ref{fig:fit_ell_log}, it is clear that the new mixing length computed from the DNS data exhibits the expected asymptotic behaviors. For the purpose of modeling, the mixing length shall be represented by the following function
\begin{equation} \label{eq:ell_n_model}
    \ell_n^+ = \kappa y^+ \left(1 - \exp\left(-(\frac{y^+}{A^+})^{2/b} \right) \right)^b ,
\end{equation}
where the parameter $b$ controls the shape of the damping and $A^+$ denotes the non-dimensional damping length scale. The exponent of $2$ is chosen to obtain the correct asymptotic near-wall behavior $\ell_n^+ \sim (y^+)^3$ according to Eq.~(\ref{eq:D_scaling}), which is derived in Appendix~\ref{app:damping}. For a more comprehensive analysis of various damping functions and their asymptotic behaviors, the readers are referred to Appendix~\ref{app:existing_models}.

In the new model (Eq.~(\ref{eq:ell_n_model})), the constant $\kappa$ is asymptotically the inverse of the log slope of the mean velocity profile at high Reynolds numbers. For instance, Nagib et al. \cite{Nagib2008} report that $\kappa \approx 0.38$ is appropriate for channels and zero pressure gradient boundary layers of Reynolds number based on the boundary layer displacement thickness $Re_{\delta^*} \approx 10,000$. And this value is consistent with the channel DNS data at $Re_\tau = 5200$ \cite{Lee2015} in Fig.~\ref{fig:fits}, where the asymptotic slope of $\ell_n$ is found to be approximately $0.38$.

One remaining issue is to properly define the parameter $b$ that controls the shape of the near-wall damping. As shown in Fig.~\ref{fig:ell_vs_y_new}, there appears to be little variation in the shape of the new mixing length even for boundary layers with strong adverse pressure gradients, and therefore a single choice of $b$ may be suitable for all flows. Considering two DNS data with the highest $Re_\tau$ available, i.e. the channel flows with $Re_\tau = 8000$ \cite{Yamamoto2018} and $Re_\tau = 5200$ \cite{Lee2015}, the optimal choices of $b$ for these datasets are 1.1 and 0.84, respectively. In the following, $b=1$ is chosen as a compromise for all flows.
 
The damping coefficient $A^+$ is now the last remaining free parameter in the model. The damping coefficient directly controls the extent of the buffer layer, or equivalently the log intercept of the velocity profile. In fact, even with an arbitrary choice for $b$ (see Appendix~\ref{app:existing_models} for details), the resulting model can provide a decent prediction of the mean velocity as long as $A^+$ is correctly calibrated. As shown in Fig.~\ref{fig:u_vs_y}, the intercept constant appears to be the primary feature that varies between different non-equilibrium flows due to the pressure gradient. Even in equilibrium flows, it is observed that the log intercept is much less universal than the k{\'a}rm{\'a}n constant $\kappa$. Therefore, in this work, $A^+$ is proposed to vary according to specific flows instead of being specified as a universal constant.

%%%%%%%%%%%%%%%%%%%%%%%%%%%%%%%%%%%%%%%%%%%%%%%%%%%%%%%%%%%%%%
\subsection{\label{sec:PG_fitting} Modeling based on pressure-gradient parameters}
%%%%%%%%%%%%%%%%%%%%%%%%%%%%%%%%%%%%%%%%%%%%%%%%%%%%%%%%%%%%%%

In equilibrium flows, such as the channel, pipe, and zero pressure gradient boundary layer, the law of the wall has needed only small modifications to optimally fit these velocity profiles. However, in pressure-gradient boundary layers, relatively large changes in the logarithmic intercept are required (see Fig.~\ref{fig:u_vs_y}). Most of the classical mixing-length-based models for non-equilibrium flows rely on a non-dimensional pressure-gradient parameter as an input.

%%in this work, the boundary layer shape factor is proposed as a better choice for sensitizing the velocity profile to non-equilibrium effects. 

\begin{figure}
\begin{subfigure}{.5\textwidth}
  \centering
  \includegraphics[width=1\linewidth]{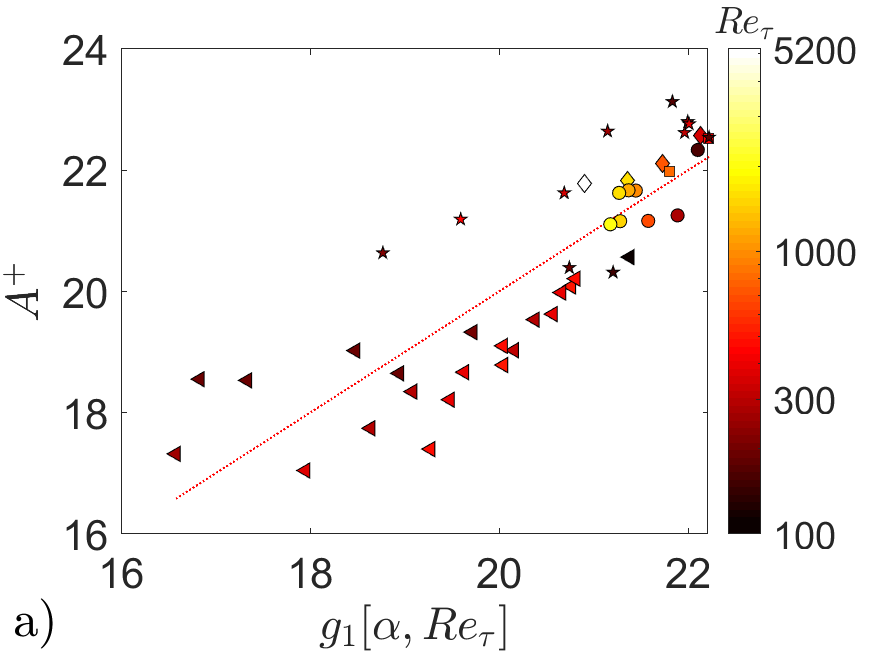}
  \captionlistentry{}
  \label{fig:A_vs_alpha}
\end{subfigure}%
\begin{subfigure}{.5\textwidth}
  \centering
  \includegraphics[width=1\linewidth]{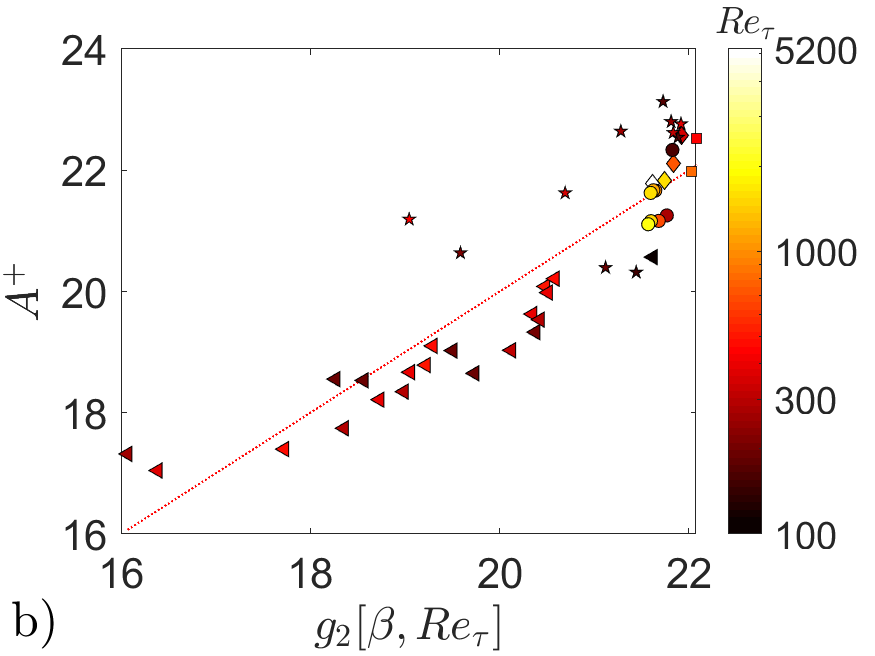}
  \captionlistentry{}
  \label{fig:A_vs_beta}
\end{subfigure}%
\caption{Distributions of the damping coefficient $A^+$ plotted versus a least-squares regression of the friction Reynolds number $Re_\tau$, and (a) the inner pressure-gradient parameter $\alpha$ or (b) the outer pressure-gradient parameter $\beta$. The red dotted lines denote the least-squares regressions characterized by the functional $g_1[\alpha,Re_\tau]=24.5-170\alpha-0.421 \ln(Re_\tau)$ (a) and $g_2[\beta,Re_\tau]=22.5-1.29\beta-0.117 \ln(Re_\tau)$ (b). The symbols denote the data from the cases in Table~\ref{tab:database}, with the symbol color indicating $Re_\tau$ and the symbol type indicating the flow type, i.e. channel flows (diamonds), ZPGBLs (circles), pipe flows (squares), APGBLs (triangles), and airfoil flows (pentagrams).}
%\label{}
\end{figure}

One popular approach for sensitizing the velocity profile to the pressure gradient considers the inner pressure-gradient parameter, i.e. the Mellor parameter, which is defined as
\begin{equation} \label{eq:define_alpha}
    \alpha = \frac{\delta_\nu}{\tau_w}\tder{P}{x},
\end{equation}
and correlates $A^+$ or the log intercept with the parameter $\alpha$ \cite{Huffman1972,Granville1989,Johnstone2010,Nickels2004,duprat2011wall}. 
Taking the cases in Table~\ref{tab:database} for instance, by modeling the mixing length $\ell_n^+$ with the model defined in Eq.~(\ref{eq:ell_n_model}), the optimal choice of $A^+$ is made to minimize the difference between the reference solution and the modeled velocity profile (the solution of Eq.~(\ref{eq:new_model_ode})) at the wall-normal distance $y=0.1\delta$. This leads to a value of $A^+$ for each profile in the database. The bilinear regression of $A^+$ versus $\alpha$ and $\ln(Re_\tau)$ is plotted in Fig.~\ref{fig:A_vs_alpha} for all cases in the database. The fitting function is shown in the caption and has a coefficient of determination $R^2 = 0.73$.

Alternatively, a better measure of non-equilibrium effects can be achieved by fitting the outer pressure-gradient parameter, i.e. the Clauser parameter, which is defined as
\begin{equation}
    \beta = \frac{\delta^*}{\tau_w}\tder{P}{x} = \frac{\delta^*}{\delta} Re_\tau \alpha,
\end{equation}
where $\delta^*$ denotes the boundary layer displacement thickness and is defined as
\begin{equation} \label{eq:define_delstar}
    \delta^* = \int_0^\delta \left( 1 - \frac{U}{U_e} \right) dy,
\end{equation}
where $U_e$ is the mean velocity at the boundary layer edge $y=\delta$.

Similar to the correlations presented above, for the given cases in Table~\ref{tab:database}, a bilinear regression of the optimal choices of $A^+$ versus $\beta$ and $\ln(Re_\tau)$ is shown in Fig.~\ref{fig:A_vs_beta}. This fit has a coefficient of determination $R^2 = 0.80$. The improved coefficient of determination suggests that $\beta$ is more indicative than $\alpha$ for determining the optimal damping coefficient $A^+$. Meanwhile, computing $\beta$ requires the definition and the computation of the boundary layer thickness $\delta$ and the edge velocity $U_e$, which are non-trivial in pressure gradient flows.

It is worth remarking that Bernard et al. \cite{Bernard2003} propose an empirical relation such that the slope of the mixing length depends on $\beta$. However, it is observed in Fig.~\ref{fig:u_vs_y} that the log slope is more universal than the log intercept constant. This motivates the present approach of correlating $A^+$, rather than $\kappa$, with $\beta$.

%%%%%%%%%%%%%%%%%%%%%%%%%%%%%%%%%%%%%%%%%%%%%%%%%%%%%%%%%%%%%%
\subsection{\label{sec:H_fitting} Modeling based on the boundary layer shape factor}
%%%%%%%%%%%%%%%%%%%%%%%%%%%%%%%%%%%%%%%%%%%%%%%%%%%%%%%%%%%%%%
As discussed above, the classical pressure-gradient parameters $\alpha$ or $\beta$ in combination with Reynolds number $Re_\tau$ do not uniquely and completely define the boundary layer velocity profile. The reason is that these pressure-gradient parameters are unaware of the spatial (or temporal in a Lagrangian sense) history of the flow, see e.g. \cite{Johnstone2010} (concerning $\alpha$) and \cite{Bobke2017} (concerning $Re_\tau$ and $\beta$). Only the local pressure gradient effect is taken into account instead of the integrated effect on the flow in the streamwise (or temporal) dimension. 

However, these integrated history effects are significant in non-equilibrium flows. The most straightforward solution for incorporating the boundary layer history effects into the wall model is to employ a PDE-based wall model, but this leads to a significantly increased computational cost. Instead, we propose that the same objective can be achieved by correlating $A^+$ with the boundary layer shape factor $H$ for the ODE-based wall model and hypothesizing that $A^+(H,Re_\tau)$ is a universal function. For boundary layers, the shape factor is defined as $H = \delta^*/\theta$, where the momentum thickness $\theta$ is computed by
\begin{equation} \label{eq:define_theta}
    \theta = \int_0^\delta \frac{U}{U_e} \left( 1 - \frac{U}{U_e} \right) dy.
\end{equation}
The rationale for this hypothesis is based on the following two observations. First, non-equilibrium effects directly modify the the boundary layer shape factor $H$ \cite{Tamaki2020}. Second, the dominant contribution to the shape factor in WMLES comes from the outer PDE solver, which captures non-equilibrium (history) effects by construction, as demonstrated in section~\ref{sec:feasibility}. These observations imply that by correlating the inner wall model with the shape factor, the history effects captured by the outer solver can be leveraged by the wall model.

\begin{figure}
\begin{subfigure}{.5\textwidth}
  \centering
  \includegraphics[width=1\linewidth]{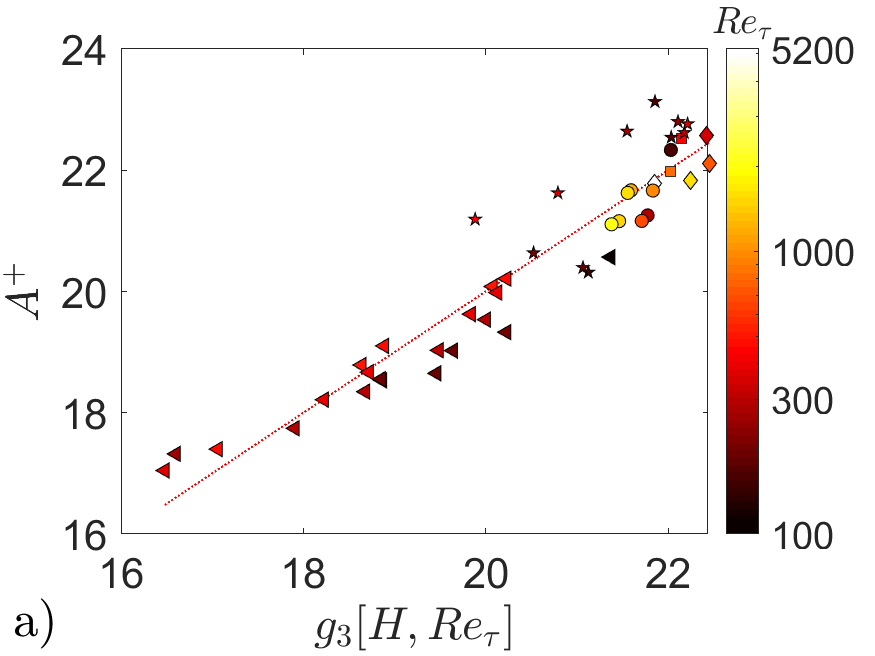}
  \captionlistentry{}
  \label{fig:A_vs_H}
\end{subfigure}%
\begin{subfigure}{.5\textwidth}
  \centering
  \includegraphics[width=1\linewidth]{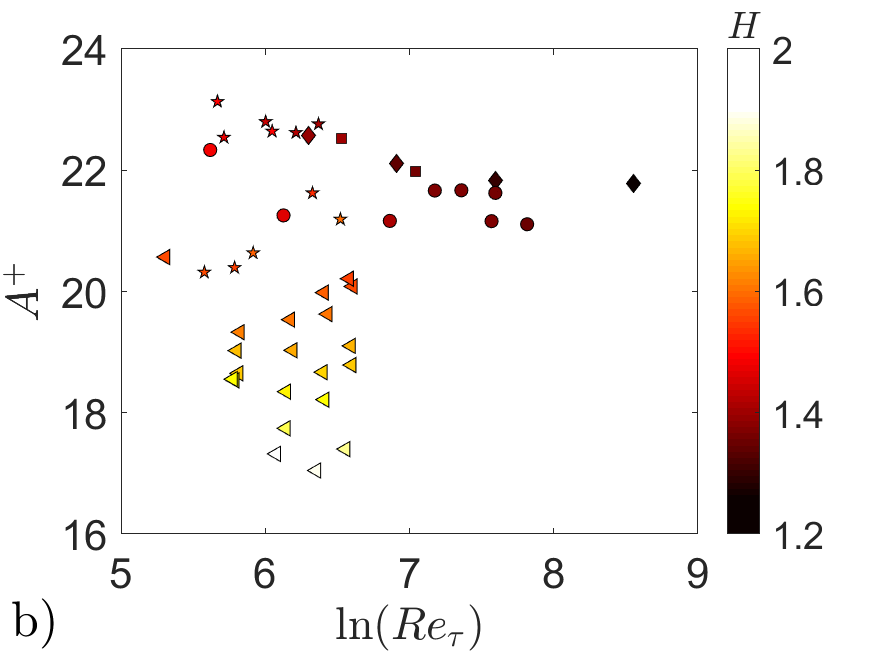}
  \captionlistentry{}
  \label{fig:A_vs_Retau}
\end{subfigure}%
\caption{Distributions of the damping coefficient $A^+$ plotted versus (a) a least-squares regression of the shape factor $H$ and $\ln(Re_\tau)$, and (b) $\ln(Re_\tau)$. In panel (a), the red dotted line denotes the least-squares regression characterized by the functional $g_3[H,Re_\tau]=45.2-11.8 H -0.993 \ln(Re_\tau)$. In both panels, the symbols denote the data from the cases in Table~\ref{tab:database}, with the symbol type indicating the flow type, i.e. channel flows (diamonds), ZPGBLs (circles), pipe flows (squares), APGBLs (triangles), and airfoil flows (pentagrams). The symbol color indicates $Re_\tau$ (a) and $H$ (b).}
%\label{}
\end{figure}

As show in Fig.~\ref{fig:A_vs_H}, with the same dataset as in Fig.~\ref{fig:A_vs_alpha} and~\ref{fig:A_vs_beta}, a much better collapse is observed when plotted versus the shape factor $H$, than versus the inner and outer pressure-gradient parameter (with a coefficient of determination $R^2 = 0.90$ compared to 0.73 and 0.80, respectively). Fig.~\ref{fig:A_vs_Retau} indicates that the present $A^+$ correlation is not merely capturing $Re_\tau$ effects.

For all three optimal linear regressions shown in Fig.~\ref{fig:A_vs_alpha},~\ref{fig:A_vs_beta}, and~\ref{fig:A_vs_H}, the coefficient for $\ln(Re_\tau)$ is negative. Consequently, $A^+\rightarrow\infty$ and thus $\nu_t \rightarrow 0$ when the Reynolds number is sufficiently low with $Re_\tau \rightarrow 0$. This is consistent with the expected behavior of the eddy viscosity that it should smoothly turn off in a laminar flow. 

It is worth noting that these regressions only apply to the present mixing length model and damping function given in Eq.~(\ref{eq:new_ell_m}), for the fully turbulent, incompressible flows with zero wall penetration. While the optimal choices of the regression coefficients may be different for other mixing-length models, the suitability of correlating $A^+$ with $H$ and $Re_\tau$ may still hold in general.

%%%%%%%%%%%%%%%%%%%%%%%%%%%%%%%%%%%%%%%%%%%%%%%%%%%%%%%%%%%%%%
\section{\label{sec:results}Performance validation of the new model}
%%%%%%%%%%%%%%%%%%%%%%%%%%%%%%%%%%%%%%%%%%%%%%%%%%%%%%%%%%%%%%

In summary, the proposed new model consists of the ODE in Eq.~(\ref{eq:new_model_ode}),  where the mixing length is parameterized as
\begin{equation} \label{eq:final_mixing_length}
    \ell_n^+ = \kappa y^+ \left(1 - \exp\left(-(\frac{y^+}{A^+[H,Re_\tau]})^{2} \right) \right) ,
\end{equation}
and the damping coefficient is given by
\begin{equation} \label{eq:final_fitting}
    A^+[H,Re_\tau] = 45.2 - 11.8 H - 0.993 \ln \left( Re_\tau \right).
\end{equation}
The computation of the shape factor $H$ will be discussed in section~\ref{sec:feasibility}. Otherwise, the ODE model is solved iteratively in the same way as for the classical models of Cabot and Prandtl. Specifically, the no-slip boundary condition is imposed at $y=0$ while the  Dirichlet boundary condition $U=U_m$ is applied at the matching location $y=y_m$, where $U_m$ is taken from the outer PDE solver. And, the matching location is typically chosen to be the first- or third- grid point of the mesh for the outer PDE solver \cite{Yang2017a,Kawai2012}. 

In this work, $y_m=0.1\delta$ is adopted as suggested in \cite{Kawai2012}. In this {\em a priori} study, the data at the matching location will be provided from DNS or WRLES, such that any resulting errors can be attributed to the wall model instead of the matching data. 
Here, the relative error $\epsilon_{\tau_w}$ is defined as the difference between the wall shear stress $\tau_w$ computed from the wall model and that from DNS or WRLES.

\begin{figure}
\begin{subfigure}{.5\textwidth}
  \centering
  \includegraphics[width=1\linewidth]{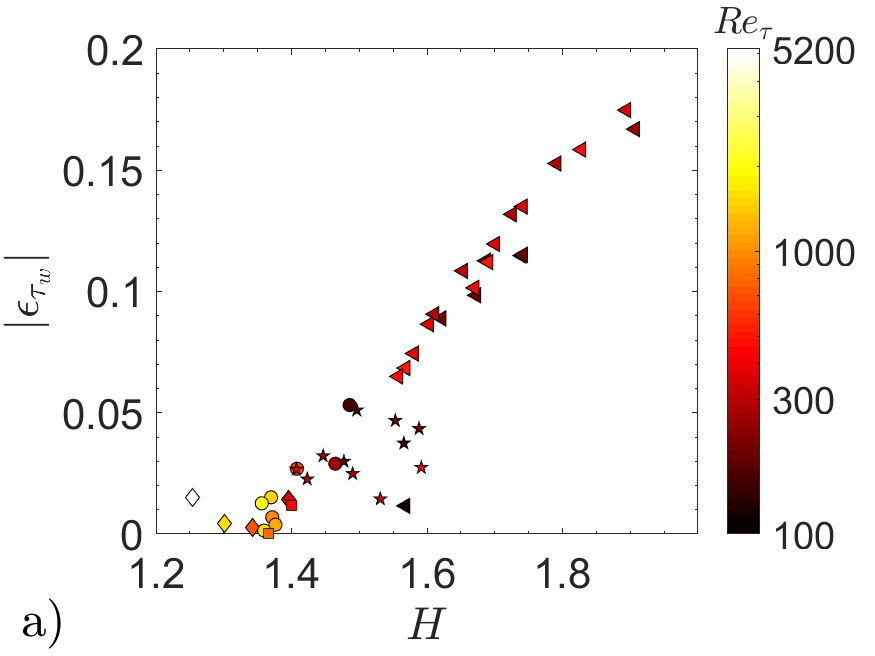}
  \captionlistentry{}
  \label{fig:err_vs_H_1}
\end{subfigure}%
\begin{subfigure}{.5\textwidth}
  \centering
  \includegraphics[width=1\linewidth]{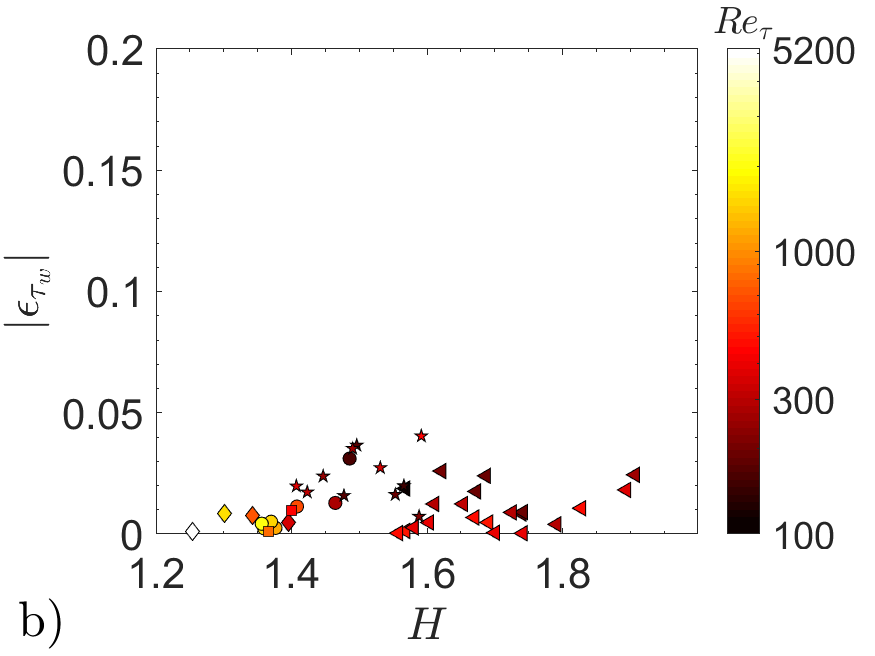}
  \captionlistentry{}
  \label{fig:err_vs_H_2}
\end{subfigure}%
\caption{Distributions of the relative error $\epsilon_{\tau_w}$ between the wall stress predicted by the well-resolved simulations and that by the wall-modeled simulations (from Cabot's model (a) and the present new model (b)) for the cases in Table~\ref{tab:database} versus the underlying shape factor $H$. In both panels, the symbols denote the data from the cases in Table~\ref{tab:database}, with the symbol color indicating $Re_\tau$ and the symbol type indicating the flow type, i.e. channel flows (diamonds), ZPGBLs (circles), pipe flows (squares), APGBLs (triangles), and airfoil flows (pentagrams).}
%\label{}
\end{figure}

\begin{figure}
\begin{subfigure}{.5\textwidth}
  \centering
  \includegraphics[width=1\linewidth]{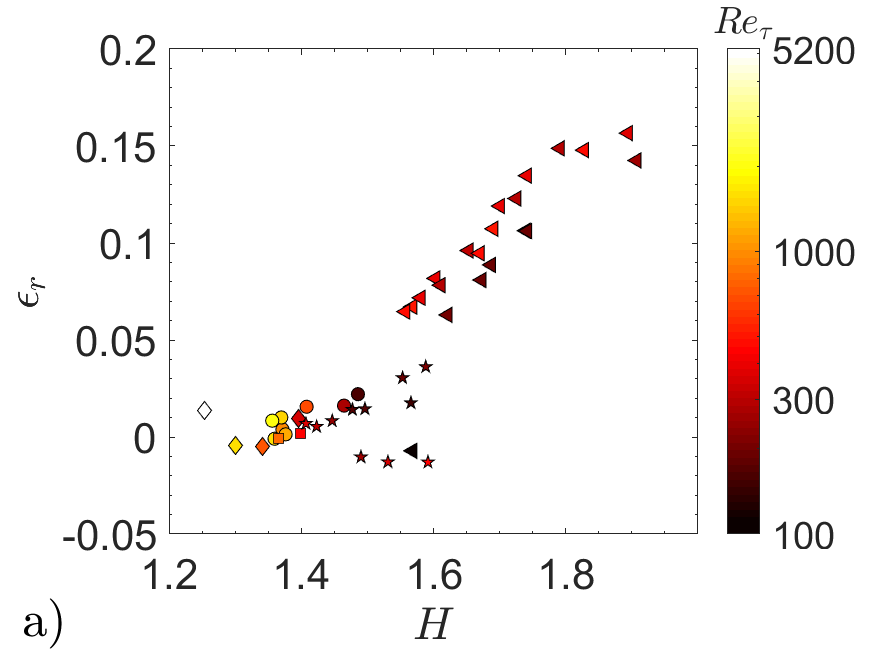}
  \captionlistentry{}
  \label{fig:err_vs_H_3}
\end{subfigure}%
\begin{subfigure}{.5\textwidth}
  \centering
  \includegraphics[width=1\linewidth]{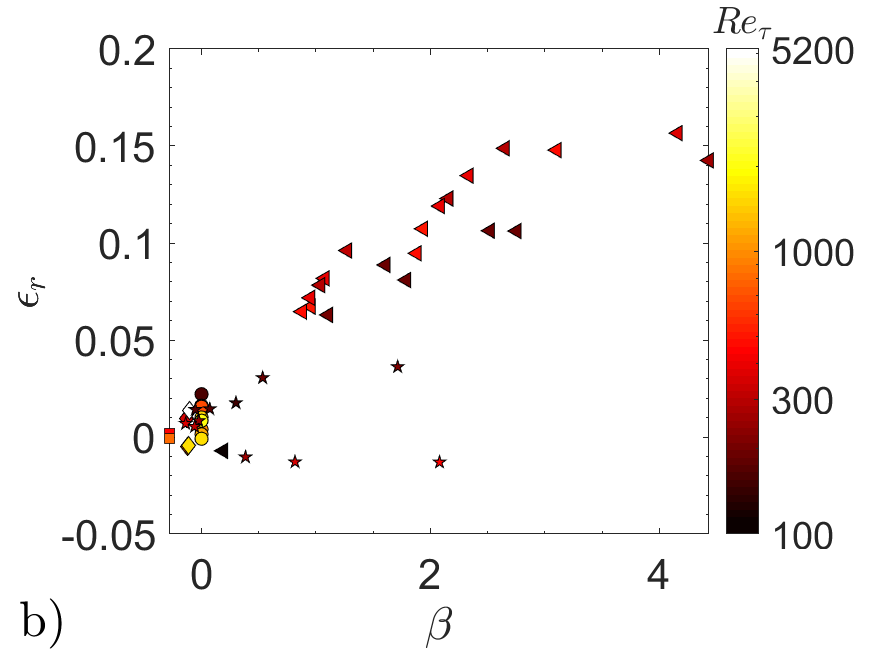}
  \captionlistentry{}
  \label{fig:err_vs_beta}
\end{subfigure}%
\caption{Distribution of the error reduction $\epsilon_r$, in terms of the wall-stress prediction by switching from Cabot's model to the new model, plotted versus the shape factor $H$ (a) and the outer pressure-gradient parameter $\beta$ (b). The symbols denote the data from the cases in Table~\ref{tab:database}, with the symbol color indicating $Re_\tau$ and the symbol type indicating the flow type, i.e. channel flows (diamonds), ZPGBLs (circles), pipe flows (squares), APGBLs (triangles), and airfoil flows (pentagrams).}
\end{figure}

As shown in Fig.~\ref{fig:err_vs_H_1}, the relative error of the wall shear stress from the classical Cabot's model is as large as $17\%$  for cases with strong pressure gradients. Meanwhile, the error from the new model, as shown in Fig.~\ref{fig:err_vs_H_2}, is typically less than $2\%$ with the maximum of $5\%$. The quantitative error reduction by deploying the new model is shown in Fig.~\ref{fig:err_vs_H_3} versus $H$ and in Fig.~\ref{fig:err_vs_beta} versus $\beta$. Cases with the strongest pressure gradients have the largest errors with the classical model and the most remarkable error reductions by switching to the new model. For very few cases, there is a tiny error increase of about $1\%$, which can be attributed to the fitting errors evident in Fig.~\ref{fig:A_vs_H}.


\begin{figure}
  \centering
  \includegraphics[width=0.5\linewidth]{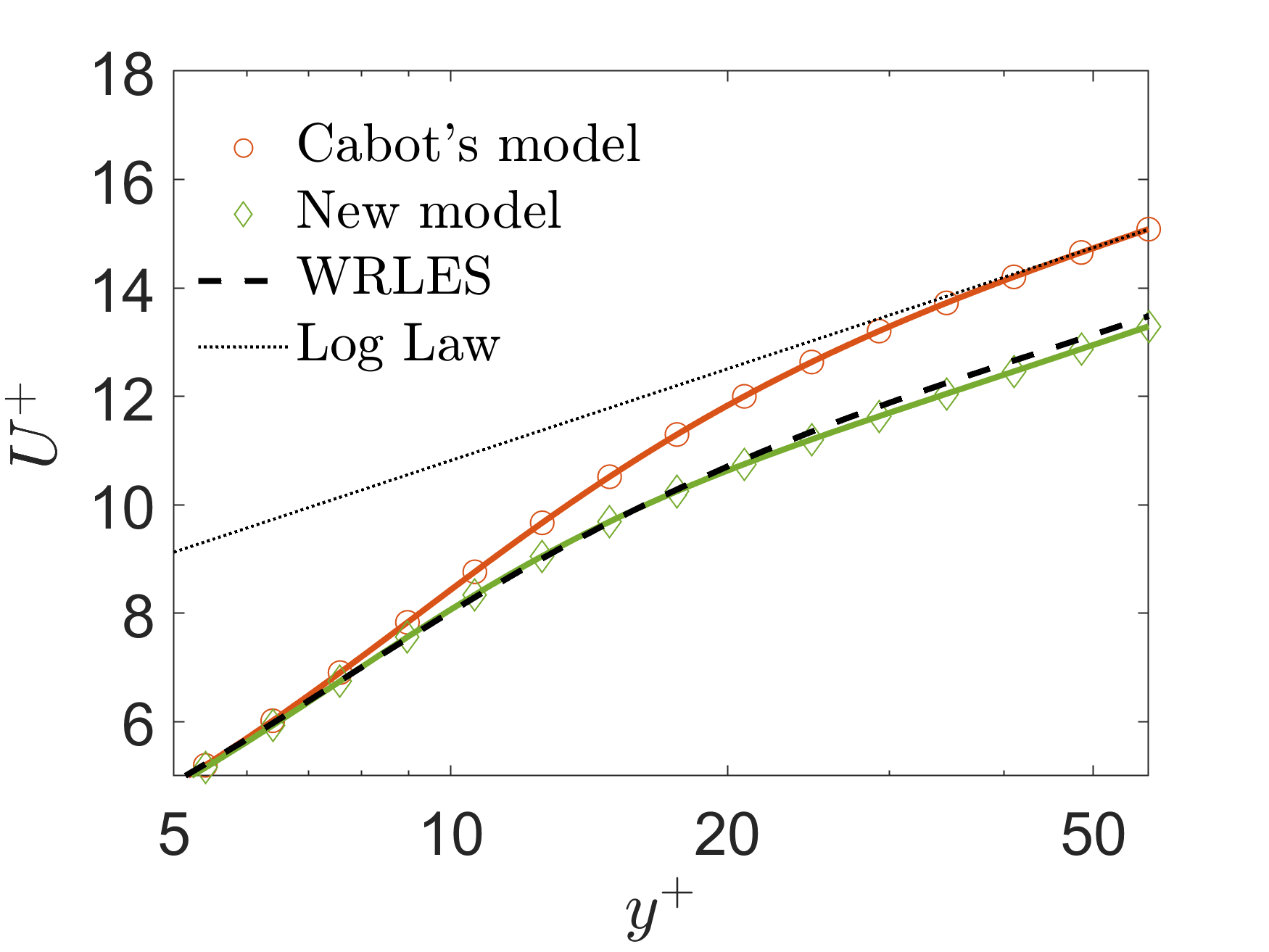}
  \caption{The streamwise velocity profile $U^+$ plotted versus the wall-normal coordinate $y^+$, for the $\beta=4.2$ case considered in Fig.~\ref{fig:u_vs_y}. Included are the WRLES prediction (dashed line), the results from the new model (green circles) and the Cabot's model (red circles), and the log law reference (dotted line).}
   \label{fig:LOW_u_vs_y_new_model}
\end{figure}

In terms of the dimensionless velocity profile, as shown in Fig.~\ref{fig:LOW_u_vs_y_new_model}, it is clear that the new model greatly improves the prediction accuracy  when compared to the classical model (Cabot's model), and the shift of the logarithmic intercept highlighted in Fig.~\ref{fig:u_vs_y} is well captured.

%%

%%%%%%%%%%%%%%%%%%%%%%%%%%%%%%%%%%%%%%%%%%%%%%%%%%%%%%%%%%%%%%
\subsection{\label{sec:feasibility}The computation of the boundary layer shape factor}
%%%%%%%%%%%%%%%%%%%%%%%%%%%%%%%%%%%%%%%%%%%%%%%%%%%%%%%%%%%%%%

One remaining critical issue of the present wall model is the estimation of the boundary layer shape factor $H$ based on the well-resolved inner ODE-based wall-model solution and the coarse outer PDE solution. The PDE solution can capture the history effects, but is under-resolved in the near-wall region. In this work, we propose to approximate the shape factor as
\begin{equation}\label{eq:H_in_out}
    H \approx \frac{\delta^*_i+\delta^*_\mathrm{o}}{\theta_i + \theta_\mathrm{o}},
\end{equation}
where $\theta$ and $\delta^*$ follow their definitions in Eq.~(\ref{eq:define_theta}) and~(\ref{eq:define_delstar}) except that the subscript $i$ refers to an integral from the wall to the location $y=y_m$, and the subscript $\mathrm{o}$ refers to an integral from $y=y_m$ to $y=\delta$. Specifically, Eq.~(\ref{eq:H_in_out}) can be written as
\begin{equation} \label{eq:H_long_form}
    H \approx \frac{\int_0^{y_m} \left( 1 - \frac{U}{U_e} \right) dy + \int_{y_m}^\delta \left( 1 - \frac{U_{\mathrm{o}}}{U_e} \right) dy}{\int_0^{y_m} \frac{U}{U_e} \left( 1 - \frac{U}{U_e} \right) dy + \int_{y_m}^\delta \frac{U_\mathrm{o}}{U_e} \left( 1 - \frac{U_\mathrm{o}}{U_e} \right) dy},
\end{equation}
where the solution $U$ is taken from the inner wall model while $U_\mathrm{o}$ is from the outer PDE solution.

Since this work does not incorporate the outer PDE simulation explicitly (in an {\em a posteriori} sense), the well-resolved simulation data from DNS or WRLES is used to compute the outer contributions to the shape factor in Eq.~(\ref{eq:H_in_out}). This permits the study of the errors inherent to Eq.~(\ref{eq:H_in_out}) in isolation from the numerical truncation errors and the subgrid-scale modeling errors in the outer solver, as these are separate issues that are not unique to the present wall model. Methods for reducing these errors will depend on the details of the outer solver, e.g. the outer solver may be RANS, LES, DES, etc.

Recalling that the proposed inner wall model for computing the velocity profile $U[y]$ explicitly depends on the parameter $H$, a two-way coupling exists between $H$ and $U[y]$. This implies that these quantities must be computed iteratively. The procedure begins with an initial guess of $H \approx \delta^*_\mathrm{o}/\theta_\mathrm{o}$ based on the outer solution. This estimate of $H$ is then fed into the ODE model for solving $U[y]$ according to Eq.~(\ref{eq:new_model_ode}),~(\ref{eq:final_mixing_length}), and~(\ref{eq:final_fitting}). As the first iteration, the resulting inner profile $U[y]$ can be used to update the estimate of $H$ according to Eq.~(\ref{eq:H_long_form}). Successive iterations involve recomputing $U[y]$ with the latest value of $H$ and recomputing $H$ based on the updated $U[y]$ profile.

\begin{figure}
\begin{subfigure}{.5\textwidth}
  \centering
  \includegraphics[width=1\linewidth]{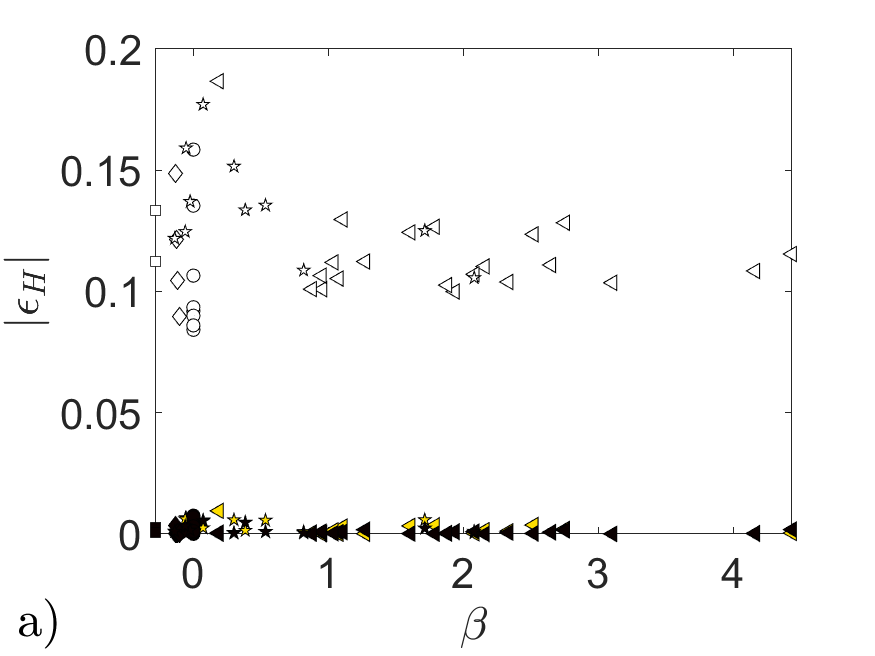}
  \captionlistentry{}
  \label{fig:H_err_vs_beta_iter}
\end{subfigure}%
\begin{subfigure}{.5\textwidth}
  \centering
  \includegraphics[width=1\linewidth]{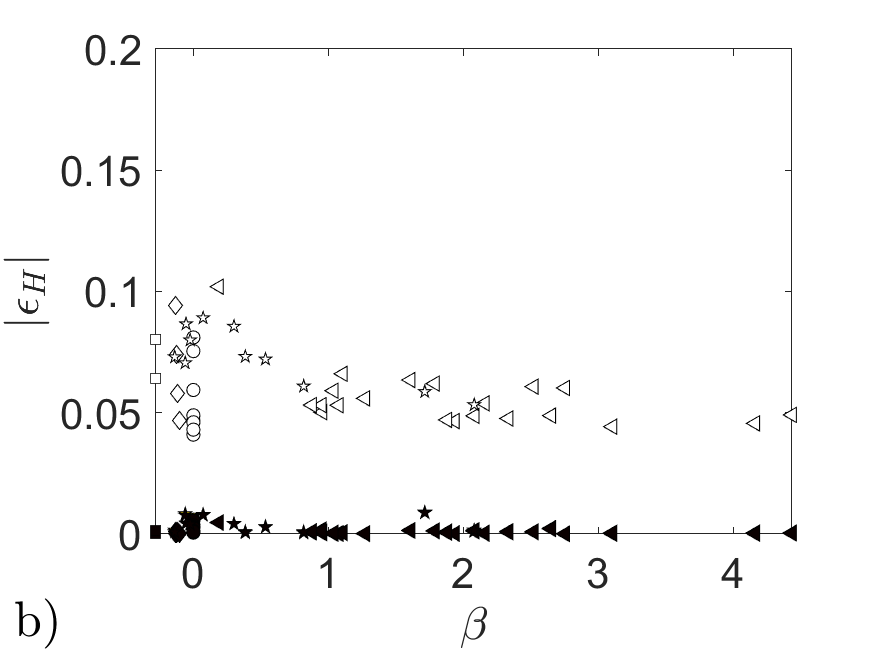}
  \captionlistentry{}
  \label{fig:H_err_vs_beta_iter_ym025}
\end{subfigure}%
\caption{Distributions of the relative error $\epsilon_H$ between the computed and the exact shape factor $H$ plotted versus the outer pressure-gradient parameter $\beta$. In both panels, the symbols denote the data from the cases in Table~\ref{tab:database}, with the symbol type indicating the flow type, i.e. channel flows (diamonds), ZPGBLs (circles), pipe flows (squares), APGBLs (triangles), and airfoil flows (pentagrams). The symbol color indicates the results computed from different iterations: the initial guess (white), the first iteration (yellow), the second (red) and tenth (black) iteration. The yellow and red symbols are obscured by the black symbols. The matching location is $y_m=0.1\delta$ (a) and $y_m=0.025\delta$ (b).}
%\label{}
\end{figure}

Fig.~\ref{fig:H_err_vs_beta_iter} shows the $H$ estimates from the iterative procedure above with the matching location $y_m = 0.1 \delta$. The initial guess for $H$ from the outer profile results in about $12\%$ error. After the first iteration, the error decays to be negligible, i.e. less than $1\%$. Successive iterations reveal that the result converges within two iterations. This rapid convergence indicates that the shape factor estimate is not sensitive to the damping coefficient $A^+$, which, on the other hand, has a large effect on the wall shear stress prediction.

As the matching location approaches the wall, the contribution to the shape factor from the inner profile vanishes. Fig.~\ref{fig:H_err_vs_beta_iter_ym025} shows the results for the case with the matching location $y_m = 0.025 \delta$ (corresponding to a fine WMLES setup with 40 points across the boundary layer \cite{Kawai2012,goc2020wall}). It is observed that the initial guess with $H \approx \delta^*_\mathrm{o}/\theta_\mathrm{o}$ still leads to a significant error, up to $10\%$. This suggests that the contribution of the inner velocity profile to the shape factor can not be neglected. Meanwhile, the proposed iterative approach approximately converges with one iteration. 

The robustness of the shape factor to the damping coefficient is because $H$ depends on the velocity profile in outer units ($U_e$ and $\delta$) rather than inner units ($u_\tau$ and $\delta_v$). Moreover, the inner wall model is constrained to satisfy the no-slip boundary condition at $y=0$ and the Dirichlet boundary condition $U(y_m)=U_\mathrm{o}(y_m)$ at the matching location $y=y_m$. These hard constraints render the $H$ estimate from the inner model contribution relatively insensitive to $A^+$. Nonetheless, the present variable $A^+$ model is required for an accurate prediction of the wall stress, which is essential for the outer solver to deliver accurate solutions.

As observed by Bobke et al. \cite{Bobke2017}, the flow history effects are required to fully characterize the flow state. Although the integrated flow history effects in the streamwise direction can be captured by the outer PDE solver in WMLES, classical wall models have not leveraged this information directly in the wall model. On the other hand, the history of the boundary layer enters the present new model explicitly through its dependence on $H$.

%%%

%%%%%%%%%%%%%%%%%%%%%%%%%%%%%%%%%%%%%%%%%%%%%%%%%%%%%%%%%%%%%%
\section{\label{sec:conclusion}Conclusions}
%%%%%%%%%%%%%%%%%%%%%%%%%%%%%%%%%%%%%%%%%%%%%%%%%%%%%%%%%%%%%%

The classical equilibrium wall model is popular since it is simple to implement in practical applications, and the performance is, in general, satisfactory for high-Reynolds-number wall-bounded turbulence. However, the prediction capability is limited due to the fact that the damping coefficient $A^+$ does not depend on the flow state. As a result, upon integration, the classical model predicts the same logarithmic intercept even in the presence of strong pressure gradients and low Reynolds numbers. Specifically, the classical models have invoked the constant-stress-layer assumption or developed approximate correlations of the shear stress profile, without making a corresponding adjustment to the eddy viscosity model, to maintain a log law. These choices are in conflict with a wide range of high-fidelity simulation data. The present model is constructed to recover the log law without the need for assumptions about or approximations of the shear stress profile.

On the other hand, while most classical stress-based wall models assume a universal value for the mixing length damping coefficient $A^+$, the new method correlates $A^+$ with the boundary layer shape factor $H$ and the friction Reynolds number $Re_\tau$. The proposed correlation of  $A^+[H,Re_\tau]$ makes a substantial improvement to the prediction of the velocity profile and the wall shear stress for a large range of Reynolds numbers and pressure gradient conditions.

The ODE-based inner model is designed to live in symbiosis with the outer PDE-based solver, which computes the velocity profile in the outer portion of the boundary layer. By feeding these data to the inner wall model, the shape factor and the wall shear stress can be accurately predicted. As a result, the new model incorporates an integral measure of the streamwise and temporal history of the flow and is accurate in non-equilibrium scenarios, while retaining similar computational efficiency as classical equilibrium models.

%%%%%%%%%%%%%%%%%%%%%%%%%%%%%%%%%%%%%%%%%%%%%%%%%%%%%%%%%%%%%%
\begin{acknowledgments}
KG acknowledges support from the National Defense Science and Engineering Graduate Fellowship and the Stanford Graduate Fellowship. LF is funded by the AFOSR Hypersonics (Grant NO. FA9550-16-1-0319). We wish to acknowledge helpful feedback from P. Moin and W. H. Ronald Chan. 
\end{acknowledgments}
%%%%%%%%%%%%%%%%%%%%%%%%%%%%%%%%%%%%%%%%%%%%%%%%%%%%%%%%%%%%%%

\appendix

%%%%%%%%%%%%%%%%%%%%%%%%
\section{Galbraith's model and the log law} \label{app:galbriath}
%%%%%%%%%%%%%%%%%%%%%%%%

Galbraith et al. \cite{Galbraith1977} employ the improved linear stress profile given in Eq.~(\ref{eq:linear_tau}) instead of invoking the constant stress layer assumption. In light of this choice, Prandtl's model, i.e. Eq.~(\ref{eq:pr_model}), is modified so that it recovers the log law even though the stress profile is not constant. The model can be defined as
\begin{equation}
    \nu_t^+ = (\ell_G^+)^2 \left| \tder{U^+}{y^+} \right|,
\end{equation}
where $\ell_G^+ = \ell_P^+ \sqrt{\tau^+}$.
Plugging this model into the velocity ODE Eq.~(\ref{eq:dudy_nondim}) leads to
\begin{equation} \label{eq:galbraith_2}
    1 = \left(\frac{1}{\tau^+} + (\ell_P^+)^2 \left|\tder{U^+}{y^+}\right| \right) \tder{U^+}{y^+}.
\end{equation}
For large wall-normal distances, the model recovers the log law, similar to Prandtl's model, while the log intercept is adapted according to the shear stress profile. For an adverse pressure gradient boundary layer, near the wall, $\tau^+ \ge 1$ (see Fig.~\ref{fig:tau_vs_y}). Consequently, when this model is deployed over the domain $0<y<0.1\delta$, the presence of $\tau^+$ in Eq.~(\ref{eq:galbraith_2}) will lead to an increase in the log intercept compared to the zero pressure gradient case where $\tau^+ \approx 1$ over this region. This adaptation is, however, opposite to that observed in Fig.~\ref{fig:u_vs_y}. Therefore, the additional complexity of including a non-constant stress profile does not result in a more predictive model.

Later, Granville \cite{Granville1989} improves the model of Galbraith et al. \cite{Galbraith1977} by sensitizing $A^+$ to the inner pressure-gradient parameter $\alpha$ (defined in Eq.~(\ref{eq:define_alpha})). The performance is similar to that plotted in Fig.~\ref{fig:A_vs_alpha}. As discussed above, sensitizing $A^+$ to the boundary layer shape factor $H$ is more robust than to the pressure-gradient parameters.

%%%%%%%%%%%%%%%%%%%%%%%%
\section{Asymptotic behavior of the damping function} \label{app:damping}
%%%%%%%%%%%%%%%%%%%%%%%%

The damping function used in Eq.~(\ref{eq:ell_n_model}) can be generalized as
\begin{equation} \label{eq:general_D_n}
    D = \left(1 - \exp\left(-(\frac{y^+}{A^+})^{n/b} \right) \right)^b ,
\end{equation}
where $n$, $b$, and $A^+$ are the model parameters that vary between models.
%%We will demonstrate that $D \sim (y^+/A^+)^n$ for small $y^+/A^+$.
By defining $z = y^+/A^+$ and $a = n/b$, the expression
\begin{equation}
    D = \left( 1 - \exp\left(-z^{a} \right) \right)^b
\end{equation}
holds.
With the series representation of the exponential term, the above equation can be further written as
\begin{equation}
    D = \left( 1 - \sum_{k=0}^{\infty} \frac{(-z^a)^k}{k!} \right)^b .
    %0 + z \tder{E}{z}\Big|_{z=0} + \frac{z^2}{2} \tder{^2 E}{z^2}\Big|_{z=0} + \mathcal{O}(z^3)
\end{equation}
By expanding the first two terms of the series, it simplifies to
\begin{equation}
    D = \left( z^a - \sum_{k=2}^{\infty} \frac{(-z^a)^k}{k!} \right)^b = \left( z^a + \mathcal{O}(z^{2a}) \right)^b ,
\end{equation}
and this implies that
\begin{equation} \label{eq:D_scaling}
    D \approx z^{n} = (y^+/A^+)^n
\end{equation}
by retaining the leading order terms.

For any proposed eddy-viscosity model, $n$ should be selected so that $\nu_t$ recovers the physical scaling of $(y^+)^3$ near the wall (see section~\ref{sec:damping} for details).

%%%%%%%%%%%%%%%%%%%%%%%%
\section{The near-wall behaviour of existing mixing-length models} \label{app:existing_models}
%%%%%%%%%%%%%%%%%%%%%%%%

There are numerous eddy-viscosity wall models in the literature \cite{Cabot1995,Galbraith1977,Piomelli1993,VANDRIEST1956,Balaras1994}, which rely on the damped linear mixing length given in Eq.~(\ref{eq:damped_mixing_length}). A large subset of these models employ an exponential damping function with the form of Eq.~(\ref{eq:general_D_n}).  In above discussions, it is shown that $n$ determines the near-wall scaling of the damping function $D$ instead of $b$ and $A^+$. 

In section~\ref{sec:damping}, it is argued that the physically correct near-wall scaling of the eddy viscosity is $\nu_t^+ \sim (y^+)^3$. For a damped linear mixing length of the form in Eq.~(\ref{eq:damped_mixing_length}), with an exponential damping function given by Eq.~(\ref{eq:general_D_n}), $\ell^+ \sim (y^+)^{n+1}$. These analyses constrain the choice of $n$ once the functional dependence of $\nu_t^+$ on $\ell^+$ is specified. For instance, Cabot's model \cite{Cabot1995} achieves the correct near-wall behaviour for $\nu_t$ by letting $n=2$ and $b=2$, $A^+ = 17$ \cite{Cabot2000}.

However, some models fail to satisfy this requirement. For the variants of Prandtl's model, $\nu_t^+ \sim (\ell^+_P)^2$. van Driest \cite{VANDRIEST1956} proposes $n=1$, $b = 1$, and $A^+ = 26$ for use with Prandtl's model; this leads to the wrong near-wall behavior since $n \neq 1/2$. Galbraith et al. \cite{Galbraith1977} employ the same damping function but multiply van Driest's non-dimensional mixing length by $\sqrt{\tau^+}$. Granville \cite{Granville1989} further improves the model by allowing $A^+$ to depend on the inner pressure-gradient parameter $\alpha$, but retains the incorrect choice of $n$.
Piomelli \cite{Piomelli1993} proposes $n=3/2$, $b=1/2$, and $A^+ = 25$ as a damping function for the subgrid-scale model (where $\nu_{t,SGS}^+ \sim D^2 \sim (y^+)^3$ for a uniform grid). Balaras and Benocci \cite{Balaras1994} recommend inserting this damping function into Prandtl's model (where $\nu_t^+ \sim (y^+ D)^2 \sim (y^+)^5$). This model has been widely used \cite{balaras1996two} and even wrongly been reported to have the correct near-wall behavior \cite{piomelli2002wall}. Although these variants of Prandtl's model feature the incorrect near-wall scaling, they provide reasonable fits of their corresponding mixing lengths away from the wall by carefully calibrating $A^+$.

Note that, the model proposed in this work uses the parameters of $n=2$, $b=1$, and $A^+=A^+[H,Re_\tau]$, which ensure the correct near-wall scaling.

%%%%%%%%%%%%%%%% Bibliography %%%%%%%%%%%%%%%%%%%%%%%%%
%\nocite{*}
\bibliography{wall_model.bib,PG_NEq_LOW.bib,incompressible_database.bib,incompressible_wall.bib}

\end{document}